# The Method of Visual Satellite Photometry

Anthony Mallama

anthony.mallama@gmail.com

**1. Summary**
Large constellations of artificial satellites are beginning to interfere with astronomical observing. Visual magnitude measurements of these spacecraft are useful for monitoring and characterizing their brightness. This paper describes the method used for recording brightness by eye.

**2. Context**
Several companies and countries have launched thousands of communication satellites into low Earth orbits. Tens times as many are expected in the coming years. Amateur and professional astronomers are concerned about their impact on the night sky (Mallama and Young 2021, and Barentine et al. 2023).

Experienced visual observers record the magnitudes of these spacecraft in order to characterize and monitor their brightness. The results are being used by satellite operators to assess how effectively their efforts to reduce satellite brightness are working.

**3. Equipment**
Satellite observation requires an optical instrument with enough light gathering power to show stars of magnitude 8. Binoculars with 50 mm objective lenses and 7 power magnification are a good choice in dark sky locations. However, larger aperture or higher power might be needed to reach magnitude 8 where there is significant light pollution or moonlight.

Binoculars with electronic image stabilization can be effective for high magnifications. Likewise, a small wide-field telescope can be mounted on a tripod for stability. The author uses 15x50 image stabilized binoculars for most of his observations.

**4. Planning observations**

The web-site https://heavens-above.com/ is a useful resource for planning. The Satellite Database feature allows the user to find observable passes for their location by searching on a spacecraft name or its NORAD number. Clicking on an individual pass opens a star map of the whole sky with the spacecraft trajectory plotted. Then, clicking on any location of the star chart gives a close-up view. Other resources such as the planetarium program, Stellarium, can also be used.

All magnitudes are useful for studying satellite brightness, but the most useful are those obtained after satellites reach their operational altitude. Orbital heights are graphed by date on J. McDowell's website, https://planet4589.org/space/con/star/stats.html. Long periods of constant height usually indicate that spacecraft are at their operational altitudes.

**5. Measuring brightness**

The brightness of a satellite is determined by comparison to reference stars in the same field of view. Figure 1 is a sky chart showing 2 stars of known magnitudes and a satellite track. As an example of brightness determination, suppose that the satellite is slightly more luminous than star B whose magnitude is 7.8 but distinctly fainter than star A of magnitude 7.0. Interpolation indicates that the magnitude of the satellite is about 7.6.

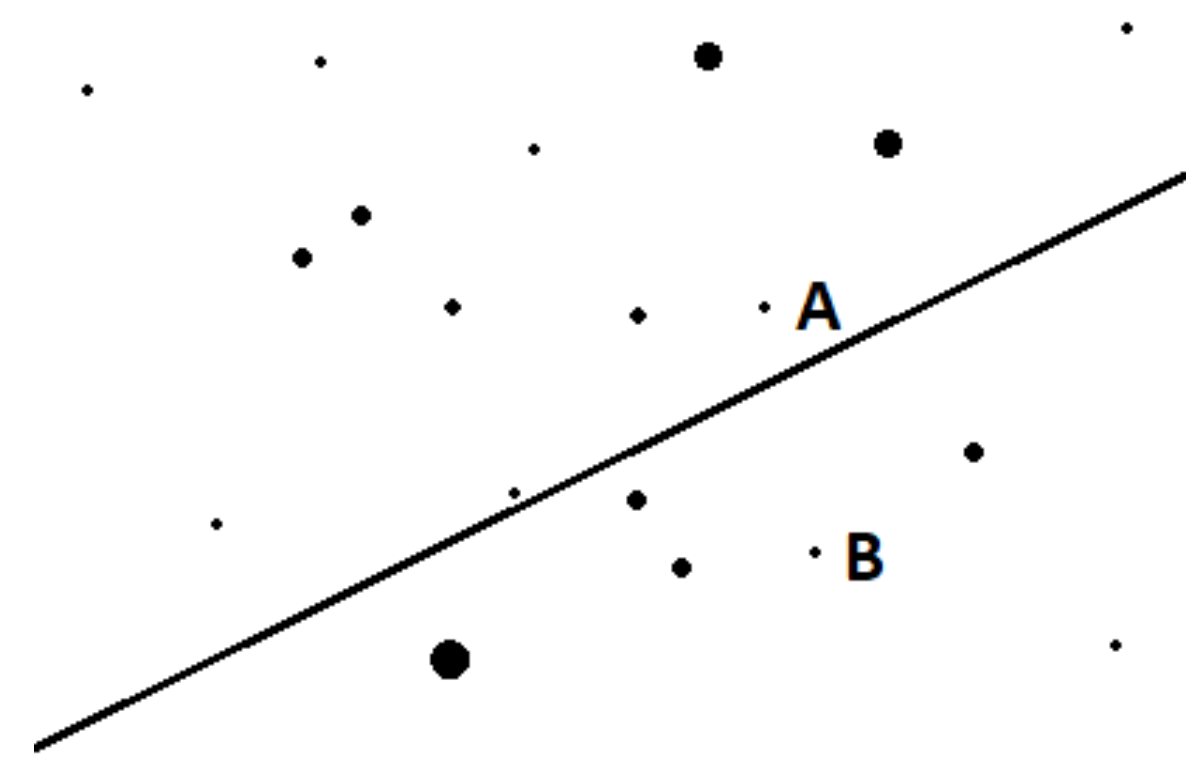

*Figure 1. This chart identifies reference stars A and B whose magnitudes are known, and shows the track of a satellite. The stars and the satellite track are fictitious.*

Reference star magnitudes can be displayed by Stellarium and other applications. The goal of visual photometry is to record satellite brightness with an accuracy of 0.2 magnitudes.

**6. Recording results**

The information needed for analysis of satellite magnitudes includes the geodetic coordinates of the observer along with the following data for each observation: the NORAD number of the satellite, the UT date and time, the satellite elevation and azimuth, the observed magnitude, the height of the satellite above the Earth, and the solar elevation and azimuth. Coordinates of the Sun, the satellite and the observer are required for investigating the effect of Sun-satellite-observer geometry on brightness. When a satellite is too faint to see, that result must also be noted. An example of this information entered on a spreadsheet is shown in Figure 2.

| Anthony Mallama | | | | | | | | |
|---|---|---|---|---|---|---|---|---|
| Coords | 38.989 | -76.769 | 49 m | | | | | |
| | | | | | | | | |
| NORAD # | UTC Date | UTC Time | Satellite | Satellite | Magnitude | Height | Solar | Solar |
| | | | Elevation | Azimuth | | km | Elevation | Azimuth |
| 60000 | 2025-Apr-25 | 1:25:58 | 65 | 239 | 7.3 | 1070 | -15 | 271 |
| 59999 | 2025-Apr-25 | 1:28:15 | 69 | 255 | 6.6 | 1065 | -15 | 272 |
| 60001 | 2025-Apr-25 | 1:30:33 | 71 | 301 | > 8.2 | 998 | -16 | 273 |
| | | | | | | | | |
| Note 1: | The entries are ficticious. | | | | | | | |
| Note 2: | The last row is an example where the satellite was too faint to be seen, so the | | | | | | | |
| | limiting magnitude is listed along with the '>' symbol. | | | | | | | |

*Figure 2. This is a sample spreadsheet containing the information needed for the study of satellite brightness.*

**7. Some previous findings**

The original Starlink spacecraft were very bright, having a mean apparent magnitude of 4.63 (Mallama 2020). After SpaceX added sunshades to the VisorSat model they were 70% dimmer (Mallama 2021). The brightness of Generation 2 Mini satellites was less than that of Gen 1 despite being four times as large (Mallama et al. 2023).

Observations of the Chinese constellations, Qianfan and Guowang, indicate mean apparent magnitudes around 6. However, these spacecraft are orbiting at 1,000 and above. Future orbits are expected to be near 500 and 300 km, which would make the satellites 1 to 2 magnitudes more luminous unless brightness mitigations are applied.

Observations of the large BlueBird satellites operated by AST SpaceMobile indicate that it is the brightest constellation with a mean apparent magnitude of 3.4 (Cole et al. 2025). The next generation of these satellites will be three times as large, which suggests that their mean magnitude will be around 2.2.

**Endnotes**

The procedures described in this paper were originally outlined in How to Estimate Satellite Magnitudes on the SeeSat-L website. More information about satellite observing techniques can be found on the Visual Satellite Observer's Home Page. The method of magnitude determination is adapted from the AAVSO Visual Observing Manual.

**Further reading**